\documentclass[12pt]{article}
\usepackage{amsmath,amssymb}
\usepackage{bm}
\usepackage{color}
\usepackage{soul}
\usepackage{graphicx}
\usepackage{cite}
\usepackage{wrapfig}
\usepackage{ulem}

\begin{document}

\title{
	\begin{flushright}
		\ \\*[-80pt]
		\begin{minipage}{0.2\linewidth}
			\normalsize
			%arXiv:YYMM.NNNN \\
			EPHOU-19-008 \\
			HUPD1907 \\*[50pt]
		\end{minipage}
	\end{flushright}
	{\Large \bf
		Modular $S_3$ invariant flavor model in SU(5) GUT
		\\*[20pt]}}

\author{
	Tatsuo Kobayashi $^{1}$
	%\footnote{A's mail}
	,~Yusuke Shimizu $^{2}$
	%\footnote{B's mail}
	,~Kenta Takagi $^{2}$
	%\footnote{C's mail}
	,\\Morimitsu Tanimoto $^{3}$
	%\footnote{D's mail}
	,~Takuya H. Tatsuishi  $^{1}$
	%\footnote{D's mail}
	\\*[20pt]
	\centerline{
		\begin{minipage}{\linewidth}
			\begin{center}
				$^1${\it \normalsize
					Department of Physics, Hokkaido University, Sapporo 060-0810, Japan} \\*[5pt]
				$^2${\it \normalsize
					Graduate School of Science, Hiroshima University, Higashi-Hiroshima 739-8526}\\*[5pt]
				$^3${\it \normalsize
				Department of Physics, Niigata University, Niigata 950-2181}
			\end{center}
		\end{minipage}}
	\\*[50pt]}

\date{
	\centerline{\small \bf Abstract}
	\begin{minipage}{0.9\linewidth}
		\medskip
		\medskip
		\small
		We present a flavor model with the  $S_3$ modular invariance  in the framework of SU(5) GUT.
		The $S_3$ modular forms of weights $2$ and $4$ give the quark and lepton mass matrices with a common complex parameter, the modulus $\tau$.
		The GUT relation of down-type quarks and charged leptons is imposed by the VEV of adjoint 24-dimensional Higgs multiplet in addition to the VEVs of $5$ and $\bar 5$ Higgs multiples of SU(5).
		The observed CKM and PMNS mixing parameters as well as the mass eigenvalues are reproduced properly.
		We discuss  the leptonic CP phase and the effective mass of the neutrinoless double beta decay with the sum of neutrino masses.
	\end{minipage}
}

\begin{titlepage}
	\maketitle
	\thispagestyle{empty}
\end{titlepage}
\newpage

% --------------------------------------------------------------------- %
% ------------------------ INTRODUCTION ------------------------------ %
% --------------------------------------------------------------------- %

\section{Introduction}
The standard model (SM) was well established by the discovery of the Higgs boson.
The SM, however, does not answer a fundamental question about the origin of flavor structure.
In order to understand the origin of the flavor structure,
many works have addressed to the discrete groups for flavors.
The  $S_3$ group was used in early models of quark masses and mixing angles
\cite{Pakvasa:1977in,Wilczek:1977uh}.
This group was also studied to explain the large mixing angle \cite{Fukugita:1998vn} in the oscillation of atmospheric neutrinos \cite{Fukuda:1998mi}. 
After the discovery of the neutrino oscillations, the discrete symmetries of flavors have been developed to reproduce observed lepton mixing angles
\cite{Altarelli:2010gt,Ishimori:2010au,Ishimori:2012zz,Hernandez:2012ra,King:2013eh,King:2014nza,Tanimoto:2015nfa,King:2017guk,Petcov:2017ggy}.
%Many models have been proposed by using 
%the non-Abelian discrete groups  $S_3$, $A_4$, $S_4$, $A_5$ and other groups with %larger orders to explain the large neutrino mixing angles.

Superstring theory with certain compactifications can lead to non-Abelian discrete flavor symmetries.
(See, e.g., \cite{Kobayashi:2006wq,Kobayashi:2004ya,Ko:2007dz,Beye:2014nxa,Olguin-Trejo:2018wpw,Nilles:2018wex,Abe:2009vi}.)
The torus and orbifold compactifications have the modular symmetry of the modulus parameter.
The flavors of both quarks and leptons transform non-trivially under the modular transformation
\cite{Lauer:1989ax,Lerche:1989cs,Ferrara:1989qb,Cremades:2004wa,Kobayashi:2017dyu,Kobayashi:2018rad}.
In this sense, the modular symmetry is a non-Abelian discrete flavor symmetry.
Yukawa couplings as well as other couplings depend on the moduli parameters in four-dimensional low-energy effective field theory derived from superstring theory.
Each coupling therefore transforms non-trivially under the modular symmetry, which is an important difference from the conventional flavor symmetries.

The modular group includes $S_3$, $A_4$, $S_4$, and $A_5$ as its finite subgroups \cite{deAdelhartToorop:2011re}.
An attractive flavor model has been put forward based on the $\Gamma_3 \simeq A_4$ modular group \cite{Feruglio:2017spp}. 
This work stimulates model buildings based on $\Gamma_2 \simeq S_3$ \cite{Kobayashi:2018vbk},
$\Gamma_4 \simeq S_4$ \cite{Penedo:2018nmg} and $\Gamma_5 \simeq A_5$ \cite{Novichkov:2018nkm}.
Phenomenological discussions of the neutrino flavor mixing have been presented
based on $A_4$ \cite{Criado:2018thu,Kobayashi:2018scp}, $S_4$ \cite{Novichkov:2018ovf} and $A_5$ \cite{Ding:2019xna} modular groups.
In particular, the comprehensive analysis of the $A_4$ modular group 
has provided a distinct prediction of the neutrino mixing angles and the CP violating phase \cite{Kobayashi:2018scp}.
The applications of the modular symmetry begin to develop  in 
quark and lepton flavors.
The $A_4$ modular symmetry also has been applied to the SU(5) grand
unified theory (GUT) of  quarks and leptons  \cite{deAnda:2018ecu},
while the residual symmetry of the $A_4$ modular symmetry has been investigated phenomenologically \cite{Novichkov:2018yse}.
The modular forms for $\Delta(96)$ and $\Delta(384)$ were also constructed \cite{Kobayashi:2018bff},
and the extension of the traditional flavor group  is discussed with modular symmetries \cite{Baur:2019kwi}.
Moreover, multiple modular symmetries are proposed as the origin of flavor\cite{deMedeirosVarzielas:2019cyj}.
The modular invariance has been also studied combined with the generalized CP symmetries for theories of flavors \cite{Novichkov:2019sqv}.
The quark mass matrix has been discussed in the $S_3$ and $A_4$ modular symmetries as well\cite{Kobayashi:2018wkl,Okada:2018yrn}.
Besides mass matrices of the quarks and leptons,
related topics such as the baryon number violation \cite{Kobayashi:2018wkl}, 
the dark matter \cite{Nomura:2019jxj},
radiatively induced neutrino masses \cite{Nomura:2019yft},
and the modular symmetry anomaly \cite{Kariyazono:2019ehj}
have been discussed.

Among them, 
the unification of quark and lepton flavors based on the modular symmetry
is an important work in the standpoint of the quark-lepton unification \cite{deAnda:2018ecu,Okada:2019uoy}
since the modulus $\tau$ is common in both quarks and leptons.
In this paper, we construct $S_3$ flavor model with modular invariance in the framework of SU(5) GUT
and discuss the Dirac CP violating phases in both quark and lepton sectors as well as the neutrino masses and mixing, the effective neutrino mass of the neutrinoless double beta decay, and Majorana CP violating phases.
We consider a six-dimensional compact space $X^6$ in addition to our four-dimensional spacetime in superstring theory.
Suppose that the six-dimensional compact space has some constituent spaces and they include a two-dimensional compact space $X^2$.
Note that $X^2$ can have geometrical symmetry such as the modular symmetry.
The quark mixing and lepton mixing are explained by a single flavor symmetry originated from $X^2$.
The modular forms for quark and lepton sectors are the same and determined by a common value of $\tau$ in our setup.
The other four-dimensional part of $X^6$ may contribute to an overall factor of Yukawa couplings, but not to their ratios.

We assume the $S_3$ modular symmetry for flavors of quarks and leptons since it is the minimal non-Abelian discrete symmetry.
Furthermore, we assume SU(5) GUT as a first step to build a realistic flavor model
with the modular invariance for both quarks and leptons.
It is emphasized that the vacuum expectation value (VEV) of 24-dimensional adjoint Higgs multiplet $H_{24}$ makes a difference between mass eigenvalues of down-type quarks and charged leptons.
Our mass matrices reproduce the observed Cabibbo--Kobayashi--Maskawa (CKM) and Pontecorvo--Maki--Nakagawa--Sakata (PMNS) parameters successfully.
We predict the leptonic CP violation phase and the effective mass of the neutrinoless double beta decay versus the sum of neutrino masses, respectively.  

This paper is composed as follows.
In section 2, we present our SU(5) GUT model with the finite modular symmetry 
$\Gamma_2 \simeq S_3$.
In section 3, we present numerical analyses of our model.
Section 4 is devoted to a summary.
Appendix A shows the modular forms of $S_3$ briefly
and appendix B presents relevant parameters in the lepton flavor mixing.

% --------------------------------------------------------------------- %
% ----------------------------- MODEL --------------------------------- %
% --------------------------------------------------------------------- %

\section{Quark and lepton mass matrices in SU(5) GUT}

Let us present our framework in supersymmetric (SUSY) SU(5) GUT.
Matter fields can be accommodated  in the $\bar F=5$ and $T=10$ representations as
\begin{align}
	F({\bf \bar{5}})=\begin{pmatrix}d_{1}^c\\d_{2}^c\\d_{3}^c\\e\\-\nu\end{pmatrix}_L,\quad
	T({\bf 10})=\begin{pmatrix}
	0        & u_{3}^c  & -u_{2}^c & u_{1} & d_{1} \\
	-u_{3}^c & 0        & u_{1}^c  & u_{2} & d_{2} \\
	u_{2}^c  & -u_{1}^c & 0        & u_{3} & d_{3} \\
	-u_{1}   & -u_{2}   & -u_{3}   & 0     & e^c   \\
	-d_{1}   & -d_{2}   & -d_{3}   & -e^c  & 0     
	\end{pmatrix}_L,
	%\quad N({\bf 1})=\nu_R^c,
\end{align}
where subscripts $1,2,3$ denote the quark colors, the superscript $c$ denote CP-conjugated fermions, and the flavor indices are omitted.
In addition, we introduce the right-handed neutrinos $N_i^c(i=1,2,3)$,
which are SU(5) singlets.
%%%%%%%%%%%%%%%%%%%%%%%%%%%%%%%%%%%%%%%%%%%%%%%%%%%%%%%%%%%%%%% 
We present the charge assignments of superfields for $\rm SU(5)$ gauge group, $S_3$ flavor symmetry and modular weights in Table \ref{tb:fields} where the subscript $i$ of $F_i$ and $T_i$ denotes the $i$-th family.
An adjoint representation of scalars $H_{24}$ breaks the SU(5) gauge symmetry and leads to the mass differences among quarks and charged leptons.
The electroweak breaking of the SM is realized by a $5(\bar 5)$ of Higgs, $H_5(H_{\bar 5})$ which also contribute to the fermion mass matrices.
These Higgs multiplets are listed in Table \ref{tb:fields}.
It also presents the modular forms of weights 2 and 4 we use.
%%%%%%%%%%%%%%%%%%%%%%%%%%%%%%%%%%%%%%%%%%%%%%%%%%%%%%%%%%%%%%%
\begin{table}[t]
	\centering
	\begin{tabular}{|c||c|c|c|c|c|c||c|c|c||c|c|} \hline
		\rule[14pt]{0pt}{0pt}
		       & $T_{1,2}$ & $T_3$ & $F_{1,2}$ & $F_3$     & $N^c_{1,2}$ & $N^c_3$ & $H_5$ & $H_{\bar{5}}$ & $H_{24}$ & $Y_{\bf 2}^{(2)}$ & $Y^{(4)}_{\bf 1}, Y^{(4)}_{\bf 2}$ \\ \hline \hline
		\rule[14pt]{0pt}{0pt}
		SU(5)  & $10$      & $10$  & $\bar{5}$ & $\bar{5}$ & 1           & 1       & $5$   & $\bar{5}$     & $24$     & 1                 & 1                                  \\
		$S_3$  & 2         & $1'$  & 2         & $1'$      & 1           & $1'$    & 1     & 1             & 1        & 2                 & 1,\ 2                              \\
		weight & $-2$      & $0$   & $-2$      & $0$       & 0           & 0       & $0$   & $0$           & $0$      & $2$               & $4$                                \\ \hline
	\end{tabular}
	\caption{
	The charge assignments of SU(5), $S_3$ and weight for superfields and modular forms.  The subscript $i$ of $F_i$ and $T_i$ denotes the $i$-th family.}
	\label{tb:fields}
\end{table}
%%%%%%%%%%%%%%%%%%%%%%%%%%%%%%%%%%%%%%%%%%%%%%%%%%%%%%%%%

For Yukawa interactions, the $S_3$ modular invariant superpotential is written as
\begin{align}
	w=w_{10}+w_{10,\bar{5}}+w_{\nu} \ , 
	\label{superpotential}              
\end{align}
where three terms of r.h.s.\,\,lead to the mass terms of up-type quarks, 
down-type quarks and charged leptons, and neutrinos, respectively. 
The up-type quark mass matrix is derived from $w_{10}$, which is explicitly given as:
\begin{align}
	\begin{aligned}
	w_{10} = &  (\alpha'_1 Y^{(4)}_{\bf 1} + \alpha'_2 Y^{(4)}_{\bf 2}) T_{1,2} T_{1,2} H_5 \left(1 +k'_1\frac{H_{24}}{\Lambda}\right ) +                                                                           
	\beta' Y^{(2)}_{\bf 2} T_{1,2}T_3 H_5 \left(1 +k'_2\frac{H_{24}}{\Lambda}\right ) \\
		&+
		%\beta'_2 Y^{(2)}_{\bf 2} T_3 T_{1,2}H_5
		%\left(1 +k'_3\frac{H_{24}}{\Lambda}\right ) +
	\gamma' T_3 T_3 H_5 \left  (1 + k'_3\frac{H_{24}}{\Lambda}\right ),
	\end{aligned}
	\label{superpotential10}
\end{align}
where $\alpha'_{1,2}$, $\beta'$, $k'_{1,2,3}$ and $\gamma'$
are  dimensionless complex constants. 
Here, $\Lambda$ denotes the cut-off scale around the SU(5) energy scale.
We set $\langle H_{24} \rangle /\Lambda =0.3$.
Thus, the next order corrections of $\langle H_{24} \rangle^2 /\Lambda^2$ are $ {\cal O}(0.1)$.
We neglect its effect because the experimental values of masses and mixing angles for the quarks and leptons include errors in ${\cal O}(10 \%)$.
We focus on the parameter regions $|k'_i| = [0,1.5]$ in the following numerical analysis.
%\begin{align}
%	w_{10} =&  (\alpha^u_1 Y^{(4){\bf 1}} + \alpha^u_2 Y^{(4){\bf 2}}) T_{1,2} T_{1,2} H_5 +
%	\beta^u_1 Y^{(2)} T_{1,2} T_3 H_5 +
%	\beta^u_2 Y^{(2)} T_3 T_{1,2} H_5 +
%	\gamma^u T_3 T_3 H_5.
%\end{align}
By using the $S_3$ tensor product of doublets  in Appendix A,
the mass matrix of up-type quarks is given in terms of modular forms $Y_1(\tau)$ 
and $Y_2(\tau)$ of Appendix A as
\begin{equation}
	M_u=
	\left(
	\begin{array}{ccc}
		\varepsilon^u & 2c'^uY_1Y_2                       & c^u_{13}Y_2  \\
		2c'^uY_1Y_2   & \varepsilon^u-2c'^u(Y_1^2 -Y_2^2) & -c^u_{13}Y_1 \\
		c^u_{13}Y_2   & -c^u_{13}Y_1                      & c^u_{33}     
	\end{array}\right),
	\label{Mu}
\end{equation}
where  the argument $\tau$ of modular forms is omitted,
and  parameters are redefined as follows:
\begin{align}
	\begin{aligned}
	\varepsilon^u & \equiv v_u [\alpha'_1(Y_1^2 + Y_2^2) + \alpha'_2(Y_1^2 - Y_2^2)](1+k'_1 \langle H_{24} \rangle /\Lambda), \\
	c'^u          & \equiv v_u\alpha'_2 (1+k'_1\langle H_{24} \rangle  /\Lambda), \quad                                       
	c^u_{13} \equiv v_u\beta' (1+k'_2 \langle H_{24} \rangle /\Lambda), 
%c^u_{31}&\equiv v_u\beta'_2(1+k'_3\langle H_{24}\rangle/\Lambda), \quad 
\quad	c^u_{33} \equiv v_u\gamma' (1+k'_3 \langle H_{24} \rangle /\Lambda),
	\label{eq:red}
	\end{aligned}
\end{align}
and  $v_u$ is the VEV for the doublet component $H_u $ of $H_5$.
This mass matrix was investigated in our previous work \cite{Kobayashi:2018wkl}.

%%%%%%%%%%%%%%%%%%%%%%%%%%%%%%%%%%%%%
Suppose the neutrinos to be Majorana particles,
which are realized by the seesaw mechanism.
Then, the neutrino mass matrix is derived from 
the superpotential  $w_{\nu}$:
\begin{align}
&	w_{\nu} =\tilde m_{3} N^c_3 N^c_3+\sum_{i=1}^2\tilde m_i N^c_i N^c_i  
	+b^\nu_3  N^c_3 F_3 H_5 \nonumber\\
&	+ a^\nu_3(Y_1F_2-Y_2 F_1)H_5N_3^c 
	+\sum_{i=1}^2 a_i^\nu(Y_1F_1+Y_2 F_2)H_5N_i^c +\Delta w_{\nu} ,          
\end{align}
where  $a^\nu_i(i=1$-$3)$, $b^\nu_3$  are  dimensionless complex constants.
The additional term  $\Delta w_{\nu}$ is the contribution to the right-handed Majorana mass terms from the dimension-five operators as:
\begin{align}
\Delta w_{\nu} =  f_3\frac{1}{\Lambda} H_{24} H_{24} N^c_3 N^c_3 +
\sum_{i=1}^2 f_i\frac{1}{\Lambda} H_{24} H_{24}  N^c_i N^c_i\,  ,
\end{align}
where $f_i$ are arbitrary coefficients.
Here, we take the diagonal basis of $N^c_i N^c_i$.

%This  superpotential is rewritten as:
%\begin{align}
%w_{\nu} =
%&\begin{aligned}
% m_3 \left [N^c_3 +\frac{1}{2m_3}(b_3^\nu L_3 H_u)\right ]^2
%-\frac{1}{4m_3}\left [b_3^\nu L_3 +a_3^\nu(Y_1 L_2-Y_2L_1)H_u\right ]^2
%\end{aligned}\nonumber \\
%&\begin{aligned} 
%+\sum_{i=1}^2 \left \{m_i\left [N^c_i+
%\frac{1}{2m_i}a_i^\nu (Y_1L_1+Y_2 L_2)H_u\right ]^2 
%-\frac{1}{4m_i} \left [a_i^\nu (Y_1 L_1 +Y_2 L_2)H_u \right ]^2  \right \}\ ,
%\end{aligned}
%\end{align}
%where the chiral superfields $L_i$
%and the SM doublet Higgs fields $H_u$ are explicitly presented. 
After integrating out $N_i^c(i=1$-$3)$ fields,
the Majorana left-handed neutrino mass matrix is therefore given as follows:
\begin{align}
	M_\nu = \frac{v_u^2}{m_{N3} }
	\begin{pmatrix}
	a_0        & 2a_2Y_1Y_2                & bY_2  \\
	2a_2Y_1Y_2 & a_0 - 2a_2(Y_1^2 - Y_2^2) & -bY_1 \\
	bY_2       & -bY_1                     & c     
	\end{pmatrix}_{LL},
	\label{numass}
\end{align}
where parameters $a_0$, $a_1$, $a_2$, $b$ and $c$ are  redefined  as:
\begin{align}
	  & \begin{aligned} 
	a_1 \equiv \frac{1}{8} \left ( \frac{m_{N3}}{m_{N1}}(a^{\nu}_1)^2+
	\frac{m_{N3}}{m_{N2}}(a^{\nu}_2)^2+(a^{\nu}_3)^2\right ) \ ,
	\qquad 
	a_2 \equiv \frac{1}{8} \left ( \frac{m_{N3}}{m_{N1}}(a^{\nu}_1)^2+
	\frac{m_{N3}}{m_{N2}}(a^{\nu}_2)^2-(a^{\nu}_3)^2\right ) \ ,
	\end{aligned} \nonumber \\
	\nonumber \\
	  & \begin{aligned} 
	b \equiv - \frac{1}{4}a^{\nu}_3 b^{\nu}_3 \ , \qquad
	c \equiv \frac{ 1}{4}(b^{\nu}_3)^2 \ , \qquad 	a_0 \equiv a_1(Y_1^2 + Y_2^2) + a_2(Y_1^2 - Y_2^2) \, ,
	\end{aligned}
\end{align}
and 
\begin{align}
 m_{Ni}\equiv \tilde m_i + f_i\frac{1}{\Lambda} \langle H_{24} \rangle^2 \,  ,
 \qquad 
 (i=1,2,3)\, .
 \label{Nmass}
\end{align}
%%%%%%%%%%%%%%%%%%%%%%%%%%%%%%%%%%%%%%%%%%%%%%%

The superpotentials for the down-type quarks and charged leptons 
are written as
\begin{align}
	\begin{aligned}
	w_{10,\bar{5}} = & (\alpha_1 Y^{(4)}_{\bf 1} + \alpha_2 Y^{(4)}_{\bf 2}) T_{1,2} F_{1,2} H_{\bar{5}} \left (1 + k_1\frac{H_{24}}{\Lambda} \right ) + 
	\beta_1 Y^{(2)}_{\bf 2}T_{1,2}F_3 H_{\bar{5}}\left(1+k_2\frac{H_{24}}{\Lambda}\right ) \\
	                 & +\beta_2Y^{(2)}_{\bf 2}T_3F_{1,2}H_{\bar{5}}\left(1+k_3\frac{H_{24}}{\Lambda}\right ) +                                           
	\gamma T_3 F_3 H_{\bar{5}} \left (1 + k_4\frac{H_{24}}{\Lambda}\right ),
	\end{aligned}
	\label{eq:down-charged}
\end{align}
where  $\alpha_{1,2}$, $\beta_{1,2}$, $k_{1,2,3,4}$ and $\gamma$
are  dimensionless complex constants.
We focus on the parameter region $|k_i| = [0,1.5]$ in the following numerical analysis.
We can construct a mass matrix for the down-type quarks and charged leptons:
\begin{align}
	M_{10,\bar{5}} = v_d
	{\footnotesize
	\begin{pmatrix}
	\alpha_0 (1 + k_1\langle H_{24} \rangle /\Lambda)         & 2\alpha_2(1 + k_1\langle H_{24} \rangle /\Lambda)Y_1Y_2                        & \beta_1(1 + k_2 \langle H_{24} \rangle/\Lambda) Y_2  \\
	2\alpha_2 (1 + k_1 \langle H_{24} \rangle /\Lambda)Y_1Y_2 & (1 + k_1 \langle H_{24} \rangle /\Lambda)[\alpha_0 - 2\alpha_2(Y_1^2 - Y_2^2)] & -\beta_1 (1 + k_2 \langle H_{24} \rangle/\Lambda)Y_1 \\
	\beta_2  (1 + k_3\langle H_{24} \rangle  /\Lambda)Y_2     & -\beta_2  (1 + k_3 \langle H_{24} \rangle/\Lambda)Y_1                          & \gamma (1 + k_4 \langle H_{24} \rangle/\Lambda)      
	\end{pmatrix}_{RL} }
	,
	\label{massmatrix10-5}
\end{align}
where we have introduced a new parameter $\alpha_0$ defined as
\begin{align}
	\alpha_0 \equiv \alpha_1(Y_1^2 + Y_2^2) + \alpha_2(Y_1^2 - Y_2^2), 
\end{align}
and  $v_d$  for the VEV of the doublet component of $H_{\bar 5}$.
We can obtain a successful mass matrix of the down-type quarks:
\begin{align}
	M_{d}=
	\begin{pmatrix}
	\varepsilon^d & 2c'^d Y_1Y_2                       & c_{13}^d Y_2  \\
	2c'^d Y_1Y_2  & \varepsilon - 2c'^d (Y_1^2 -Y_2^2) & -c_{13}^d Y_1 \\
	c_{31}^d Y_2  & -c_{31}^d Y_{1}                    & c_{33}^d      
	\end{pmatrix},
	\label{dqmatrix}
\end{align}
where we have redefined some parameters likewise with Eq.\,\eqref{eq:red}.

The quark mass matrices in Eqs.\,(\ref{Mu}) and (\ref{dqmatrix}) can reproduce the observed CKM mixing matrix elements and quark mass ratios at the GUT scale \cite{Antusch:2013jca,Bjorkeroth:2015ora}.
Indeed, we have obtained the successful up-type and down-type quark mass matrices with the hierarchical flavor structure, which are completely consistent with observed masses and CKM parameters \cite{Kobayashi:2018wkl}.
%We obtained enough parameter sets including the value of modulus $\tau$ which reproduce quark masses and CKM mixing. 
%We show a typical sample of our parameter sets:
%\begin{align}
%	\begin{aligned}
%	&{\rm Re } [\tau] = 0.42,\quad
%	&&{\rm Im }[\tau]=1.43, \\
%	&\varepsilon^u =8.44\times 10^{-6}~\mathrm{e}^{0.46\pi i} , \quad
%	&&\varepsilon^d =1.15\times 10^{-3}~\mathrm{e}^{-0.43\pi i},\\ 
%	&c'^u  = 8.37\times 10^{-3}~\mathrm{e}^{0.46\pi i}, \quad
%	&&c'^d  = 0.44~\mathrm{e}^{-0.42\pi i},  \\
%	&c^u_{13}=0.30~\mathrm{e}^{0.81\pi i}, \quad
%	&&c_{13}^d=0.23~\mathrm{e}^{0.60\pi i}, \\
%	&c^u_{31}=0.53~\mathrm{e}^{-0.61\pi i}, \quad
%	&&c_{31}^d=0.35~\mathrm{e}^{-0.66\pi i},
%	\end{aligned}
%	\label{eq:dpara}
%\end{align}
%in $c^{u}_{33} = c_{33}^d =1$ GeV units.
%For the sample value of $\tau$ in Eq.\eqref{eq:dpara}, we have
%\begin{align}
%	Y_1 = 0.125\,\mathrm{e}^{-4.65\times 10^{-3} \pi i}~, \quad
%	Y_2 = 1.96\times10^{-2} \,\mathrm{e}^{-0.42\pi i}.
%\end{align}

% --------------------------------------------------------------------- %
% --------------------------------------------------------------------- %
%%%%%%%%%%%%%%%%%%%%%%%%%%%%%%%%%%%%%%%%%%%%%%%%%%%%%%%%%%%%%%%%%%%%%%%%
%\subsection{GUT relation in SU(5) framework}
Let us discuss the charged lepton mass matrix which is possibly related to the
down-type quark mass matrix by using the SU(5) GUT relation.
We rewrite coefficients  in down-type quark mass matrix elements in terms of 
sum of contributions from VEVs of $H_{\bar 5}$ and $H_{24}$ as follows:
\begin{align}
	\varepsilon^d = \varepsilon^5 + \varepsilon^{24}, \quad 
	c'^d = c'^5 + c'^{24}, \quad                            
	c_{13}^d = c_{13}^5 + c_{13}^{24}, \quad                
	c_{31}^d = c_{31}^5 + c_{31}^{24}, \quad                
	c_{33}^d = c_{33}^5 + c_{33}^{24} \ ,                   
	\label{eq:separate}                                     
\end{align}
where we have the following relations for parameters of Eq.(\ref{massmatrix10-5}),
\begin{align}
	\alpha_0 = \varepsilon^5/v_d, \quad 
	\alpha_2 = c'^5/v_d, \quad          
	\beta_1 = c_{31}^5/v_d, \quad       
	\beta_2 = c_{13}^5/v_d, \quad       
	\gamma = c_{33}^5/v_d  \ .          
\end{align}
Let us give the Clebsch--Gordan (CG) factor $C$
which is derived by the ratio of VEVs for the charged lepton sector and down-type quark sector:
\begin{align}
	C \equiv \frac{\langle H_{24}^l \rangle}{\langle H_{24}^q\rangle} = -3/2 \ , 
\end{align}
since $H_{24}$ takes the VEV as $\langle H_{24} \rangle \propto \mathrm{diag}[2,2,2,-3,-3]$.
% \begin{align}
% 	C_1 \equiv \frac{v_d\alpha_0 k_1 H_{24}/\Lambda}{\varepsilon^{24}}, \quad
% 	C_2	\equiv \frac{v_d\beta_1 k_2 H_{24}/\Lambda}{c_{31}^{24}}, \quad
% 	C_3 \equiv \frac{v_d\beta_2 k_3 H_{24}/\Lambda}{c_{13}^{24}}, \quad
% 	C_4 \equiv \frac{v_d\gamma k_4 H_{24}/\Lambda}{c_{33}^{24}}.
% \end{align}
The charged lepton mass matrix is therefore obtained 
in terms of the elements of down-type quark mass matrix and the  coefficient $C$ by transposing the down-type quark mass matrix:
\begin{align}
	M_e = \begin{pmatrix}
	\varepsilon^5 + C \varepsilon^{24} & 2(c'^5 +C c'^{24}) Y_1 Y_2                                               & (c_{31}^5 + C c_{31}^{24}) Y_2  \\
	2(c'^5 +C c'^{24}) Y_1 Y_2         & (\varepsilon^5 + C \varepsilon^{24}) - 2(c'^5 +C c'^{24})(Y_1^2 - Y_2^2) & -(c_{31}^5 + C c_{31}^{24}) Y_1 \\
	(c_{13}^5 + C c_{13}^{24}) Y_2     & -(c_{13}^5 + C c_{13}^{24}) Y_1                                          & c_{33}^5 + C c_{33}^{24}        
	\end{pmatrix}.
	\label{emass}
\end{align}

In the quark and lepton sectors,  we obtained enough parameter sets including the value of modulus
$\tau$ which reproduce quark and lepton masses and CKM mixing. 
For example, we set 
\begin{equation}
	{\rm Re } [\tau] = 0.465,\qquad
	{\rm Im }[\tau] = 1.31, 
\end{equation}
which lead to $Y_1 = 0.116~\mathrm{exp}[4.98 \times 10^{-4} \pi i]$ and
 $Y_2 = 0.0267~\mathrm{exp}[0.461 \pi i]$.
We show a typical sample of our parameter sets:

%%%%%%%%%%%%%%%%%%%%%%%%%
\begin{align}
\begin{aligned}
&\varepsilon^u = 7.81\times 10^{-6}~\mathrm{e}^{-0.508\pi i},
&   & \varepsilon^5 = 6.42\times 10^{-4}~\mathrm{e}^{-0.788\pi i},     
&   & \varepsilon^{24} = 2.19\times 10^{-4}~\mathrm{e}^{-0.874 \pi i}, \\
&c'^u  = 2.04\times 10^{-4}~\mathrm{e}^{-0.807\pi i},
&   & c'^5  = 2.42\times 10^{-2}~\mathrm{e}^{-0.174\pi i},                          
&   & c'^{24} = 8.29\times 10^{-3}~\mathrm{e}^{-0.261\pi i},                         \\
&c^u_{13} = 0.443~\mathrm{e}^{0.802\pi i},
&   & c_{13}^5 = 2.12~\mathrm{e}^{0.184\pi i},                        
&   & c_{13}^{24} = 0.619~\mathrm{e}^{-0.876\pi i},                     \\
&
&   & c_{31}^5 = 1.03~\mathrm{e}^{-0.505\pi i},                        
&   & c_{31}^{24} = 0.428~\mathrm{e}^{0.561\pi i},                    \\
&\phantom{=}
&   & c_{33}^5 = 0.995~\mathrm{e}^{-0.0524\pi i},                       
&   & c_{33}^{24} = 0.164~\mathrm{e}^{0.465\pi i},                    
\end{aligned}
\label{eq:sample}
\end{align}
%%%%%%%%%%%%%%%%%%%%%%%%%
in $c^{u}_{33} = c_{33}^d =1$ GeV units.
These are obtained from the parameters 
%%%%%%%%%%%%%%%%%%%%%%%
\begin{align}
\begin{aligned}
&\alpha_1^\prime  = 1.05~\mathrm{e}^{0.201\pi i},
&   & \alpha_2^\prime = 1.92 \times 10^{-4}~\mathrm{e}^{-0.797\pi i}, 
&   &    \\
& \beta^\prime  = 0.417~\mathrm{e}^{0.801\pi i},
&   & \gamma^\prime   = 0.935~\mathrm{e}^{0.00331\pi i},
&   &        \\
& k_1^\prime  = 0.238~\mathrm{e}^{-0.166\pi i},
&   & k_2^\prime   = 0.210~\mathrm{e}^{0.0156\pi i},
&   & k_3^\prime  = 0.236~\mathrm{e}^{-0.0502\pi i} ,
\end{aligned}
\label{eq:original-up}
\end{align}
%%%%%%%%%%%%%%%%%%%%%%%
in $w_{10}$ of Eq.\,(\ref{superpotential10}), and  the parameters
%%%%%%%%%%%%%%%%%%%%%%%%%%%%%%%%%
\begin{align}
\begin{aligned}
&\alpha_0  = 6.42 \times 10^{-4}~\mathrm{e}^{-0.788\pi i},
&   & \alpha_2 = 0.0242~\mathrm{e}^{-0.174\pi i}, 
&   & \beta_1  = 1.03~\mathrm{e}^{-0.505\pi i},  \\
&\beta_2  = 2.12~\mathrm{e}^{0.184\pi i},
&   & \gamma   = 0.995~\mathrm{e}^{-0.0524\pi i}, 
&   & k_1  = 0.342~\mathrm{e}^{-0.0864\pi i},   \\
&k_2 = 0.292~\mathrm{e}^{0.940\pi i},
&   & k_3 = 0.416~\mathrm{e}^{-0.935\pi i},       
&   & k_4 = 0.165~\mathrm{e}^{0.517\pi i},      
\end{aligned}
\label{eq:original-down}
\end{align}
%%%%%%%%%%%%%%%%%%%%%%%%%%%%%%%%%
in $w_{10, \bar{5}}$  of Eq.\,(\ref{eq:down-charged}).

This sample parameter set leads to the following result of the CKM mixing parameters:
%%%%%%%%%%%%%%%%%%%%%%%
\begin{align}
|V_{CKM}| =
\begin{pmatrix}
0.9746 & 0.2243 & 0.0025 \\
0.2238 & 0.9745 & 0.0180 \\
0.0040 & 0.0177 & 0.9998 
\end{pmatrix},\qquad
\delta_{CP}^{CKM} = 71.18[^\circ],
	\label{eq:sample_quark1}
\end{align}
as well as the proper hierarchy of quark and charged lepton masses.
We use the above parameters for the prediction of the  neutrino sector 
in the next section.

% --------------------------------------------------------------------- %
% ---------------------- NUMERICAL RESULT---------------------------- %
% --------------------------------------------------------------------- %
%%%%%%%%%%%%%%%%%%%%%%%%%%%%%%%%%%%%%%%%%%%%%%%%%%%%%%%%%
%%%%%%%%%%%%%%%%%%%%%%%%%%%%%%%%%%%%%%%%%%%%%%%%%%%%%%%%%
\section{Numerical result}
We have obtained parameter regions that reproduce the observed fermion mass ratios and CKM  mixing parameters.
Our results are consistent with the experimental result of quark mass ratios and charged lepton mass ratios at the GUT scale within the $1\sigma$ range \cite{Antusch:2013jca,Bjorkeroth:2015ora}
%%%%%%%%%%%%%%%%%%%%%%%%%%%%%%%%%%%%%%%%%%%
\,\footnote{The quark masses are obtained at the GUT scale $2\times 10^{16}$ GeV by putting $v_u/v_d=10$ in the  minimal supersymmetric standard model,
	where the SUSY breaking scale is taken to be $1$--$10$TeV.
	In the region of $\tan\beta=3$--$10$, our numerical values are  changed 
only  by a few percent.
The proton decay may favor the larger SUSY breaking scale such as $10$TeV
as discussed in subsection 3.3.}.
%%%%%%%%%%%%%%%%%%%%%%%%%%%%%%%%%%%%%%%%%%%
In the following subsections,
we present predictions in the neutrino sector and discuss the correlation between CKM and PMNS mixing parameters.

Since we have separated the parameters of   the down-type quarks
such as $\varepsilon, c', c_{13}, c_{31}$ and $c_{33}$ into two terms
as defined in Eq.\eqref{eq:separate}, 
we can scan the parameters in the charged lepton mass matrix of Eq.(\ref{emass})
by using the successful  parameter sets of the down-type quark sector.
A typical sample is presented in Eq.(\ref{eq:sample}). 
The  parameters of the neutrino mass matrix of Eq.(\ref{numass}) have been scanned in the region of $0 < |a_0| < 2$,
$0 < |a_2| < 50$
and $0 < |b| < 15$ in $c=1$ unit
while  phases have been scanned in $[-\pi, \pi]$.
We present a sample point satisfying  the recent neutrino oscillation experimental data \cite{NuFIT} as well as the fermion mass ratio and CKM mixing parameters at the GUT 
%%%%%%%%%%%%%%%%%%%%%%%%%%%%%%%%%%%%%
\footnote{
	We have neglected the renormalization corrections for the neutrino masses and mixing parameters
	although the numerical analysis should be presented at GUT scale.
	A numerical estimation of the quantum corrections in \cite{Haba:1999fk} showed that the corrections are negligible as far as the neutrino mass scale is smaller than $200$\,meV and $\tan \beta \leq 10$.
	See also \cite{Antusch:2005gp,Criado:2018thu}.
}.
%%%%%%%%%%%%%%%%%%%%%%%%%%%%%%%%%%%%%

% --------------------------------------------------------------------- %
% --------------------------------------------------------------------- %

\subsection{Neutrino phenomenology}

In our numerical study, we have set  $\langle H_{24} \rangle /\Lambda =0.3$ with $\langle H_{24} \rangle \simeq 2 \times 10^{16}$ GeV.
Then,  we have  $\langle H_{24} \rangle^2 /\Lambda\simeq 6\times 10^{15}$GeV
which is related to the right-handed neutrino mass in Eq.(\ref{Nmass}).
Therefore, we take $m_{Ni}\simeq 10^{15}$GeV  by choosing relevant values for $\tilde m_i$ and $f_i$
\footnote{We may consider  $f_i\simeq 1/6$ or the  cancellation due to phases of $\tilde m_i$ and $f_i$.}.
Then, neutrino Yukawa couplings are found to be at most $1.3$ by inputting the experimental data.
Thus, our setup is reasonably accepted in the neutrino phenomenology if $m_{Ni}\simeq 10^{15}$GeV is taken.
We also discuss the proton decay in this setup later.

Our lepton mass matrices of Eqs.(\ref{numass}) and (\ref{emass}) reproduce the experimental result of neutrino mass squared differences and the three mixing angles within $3\sigma$ range \cite{NuFIT}.
The following results are constrained by the cosmological bound of the sum of  three light neutrino masses $m_i$, which is  $\Sigma m_i < 0.12$ eV \cite{Vagnozzi:2017ovm,Aghanim:2018eyx}.
%%%%%%%%%%%%%%%%%%%%%%%%%%%%%%%%  NH  %%%%%%%%%%%%%%%%%%%%%%%%%%%%%%
At first,
we show two sample parameter sets leading to the successful results, which are completely consistent with the observed  CKM and PMNS matrices.
We obtain a prediction for the normal hierarchy (NH) of neutrino masses from the parameter set of Eq.(\ref{eq:sample}) and the following parameters:
\footnote{
	Note that $a_2/c$ and $b/c$ are larger than ${\cal O}(1)$, but they are
	parameters and 
	couplings $a_2Y^{(4)}_{\bf 2}$ and $bY^{(2)}$ themselves are smaller than 1.}
%%%%%%%%%%%%%%%%%%%%%
\begin{align}
\frac{a_0}c = 1.61~\mathrm{e}^{0.525\pi i}, \qquad 
\frac{a_2}c = 178~\mathrm{e}^{0.502\pi i}, \qquad 
\frac{b}c = 18.0~\mathrm{e}^{-0.997\pi i},         
\end{align}
%%%%%%%%%%%%%%%%%%%%%
in which 
$c v_u^2/m_{N3}$ is a typical neutrino mass scale.
If we take the right-handed neutrino mass $m_{N3}$ to be smaller than
$10^{14}$ GeV, $c$ is less than  $0.1$.
%This sample parameter set leads to the following result of the CKM mixing parameters:
%\begin{align}
%	\begin{aligned}
%	|V_{CKM}| = \begin{pmatrix}
%	0.9739& 0.2254& 0.0028\\
%	0.2267& 0.9737& 0.0243\\
%	0.0047& 0.0240& 0.9997
%	\end{pmatrix}, \qquad
%	\delta_{CP}^{CKM} = 59.66[^\circ].
%	\end{aligned}
%	\label{eq:sample_quark1}
%\end{align}
We obtain the three lepton mixing angles $\theta_{12}$, $\theta_{23}$ and $\theta_{13}$,
the Dirac CP violating phase $\delta_{CP}$, neutrino masses, the effective mass of the neutrinoless double beta decay $\langle m_{ee} \rangle$ and the Majorana phases $\alpha_{21}$ and $\alpha_{31}$
(see notations in Appendix B) as follows:
%%%%%%%%%%%%%%%%%%%%%%%%%%%%%%%%%%%
\begin{align}
\begin{aligned}
& \sin^2\theta_{12} = 0.287, \quad                              
\sin^2\theta_{23} = 0.604, \quad
\sin^2\theta_{13} = 0.0208, \quad
\delta_{CP} = -89.6 [^\circ], \\
& \Delta m_{21}^2 = 7.14 \times 10^{-5} [\mathrm{eV}^2] ,\quad 
\Delta m_{31}^2 = 2.60 \times 10^{-3} [\mathrm{eV}^2], \quad
m_1 = 11.7 [\mathrm{meV}], \\
& \sum_i m_i = 117 [\mathrm{meV}], \quad                      
\langle m_{ee}\rangle  = 13.1 [\mathrm{meV}], \quad
\alpha_{21} = - 23.3 [^\circ], \quad
\alpha_{31} = 168 [^\circ].
\end{aligned}
\label{eq:result_NH}
\end{align}
%%%%%%%%%%%%%%%%%%%%%%%%%%%%%%%%%%%

%%%%%%%%%%%%%%%%%%%%%%%%%%%%%%%  IH  %%%%%%%%%%%%%%%%%%%%%%%%%%%%%%%%
For the inverted hierarchy (IH) of neutrino masses, 
 we use the same parameter values as the NH except
%%%%%%%%%%%%%%%%%%%%%%%%%%%%%%
\begin{align}
\frac{a_0}c = 2.57~\mathrm{e}^{0.536\pi i}, \qquad 
\frac{a_2}c = 13.6~\mathrm{e}^{0.620\pi i}, \qquad 
\frac{b}c = 6.79~\mathrm{e}^{0.313\pi i}.       
\end{align}
%which give us  CKM mixing parameters:
%\begin{align}
%\begin{aligned}
%|V_{CKM}| = \begin{pmatrix}
%0.9739& 0.2268& 0.0028\\
%0.2267& 0.9737& 0.0243\\
%0.0047& 0.0240& 0.9997
%\end{pmatrix}, \qquad
%\delta_{CP}^{CKM} = 59.66[^\circ].
%\end{aligned}
%\label{eq:sample_quark2}
%\end{align}
Then, we obtain:
%%%%%%%%%%%%%%%%%%%%%%%
\begin{align}
\begin{aligned}
& \sin^2\theta_{12} = 0.314, \quad                              
\sin^2\theta_{23} = 0.521, \quad
\sin^2\theta_{13} = 0.0244, \quad
\delta_{CP} = -90.4 [^\circ], \\
& \Delta m_{21}^2 = 7.64 \times 10^{-5} [\mathrm{eV}^2] ,\quad 
\Delta m_{31}^2 = - 2.52 \times 10^{-3} [\mathrm{eV}^2], \quad
m_3 = 11.4 [\mathrm{meV}], \\
& \sum_i m_i = 115 [\mathrm{meV}], \quad                      
\langle m_{ee}\rangle  = 47.2 [\mathrm{meV}], \quad
\alpha_{21} = 43.3 [^\circ], \quad
\alpha_{31} = 46.4 [^\circ].
\end{aligned}
\label{eq:result_IH}
\end{align}
%%%%%%%%%%%%%%%%%%%%%%%%%%%%%%%%%%%%%%%%%%%%%%%%%%%%%%%%%%%%%%%%%%%%%
% --------------------------- %
\begin{figure}[h!]
	\begin{tabular}{ccc}
		\begin{minipage}{0.475\linewidth}                                                                                               
		\includegraphics[bb=0 0 400 280,width=\linewidth]{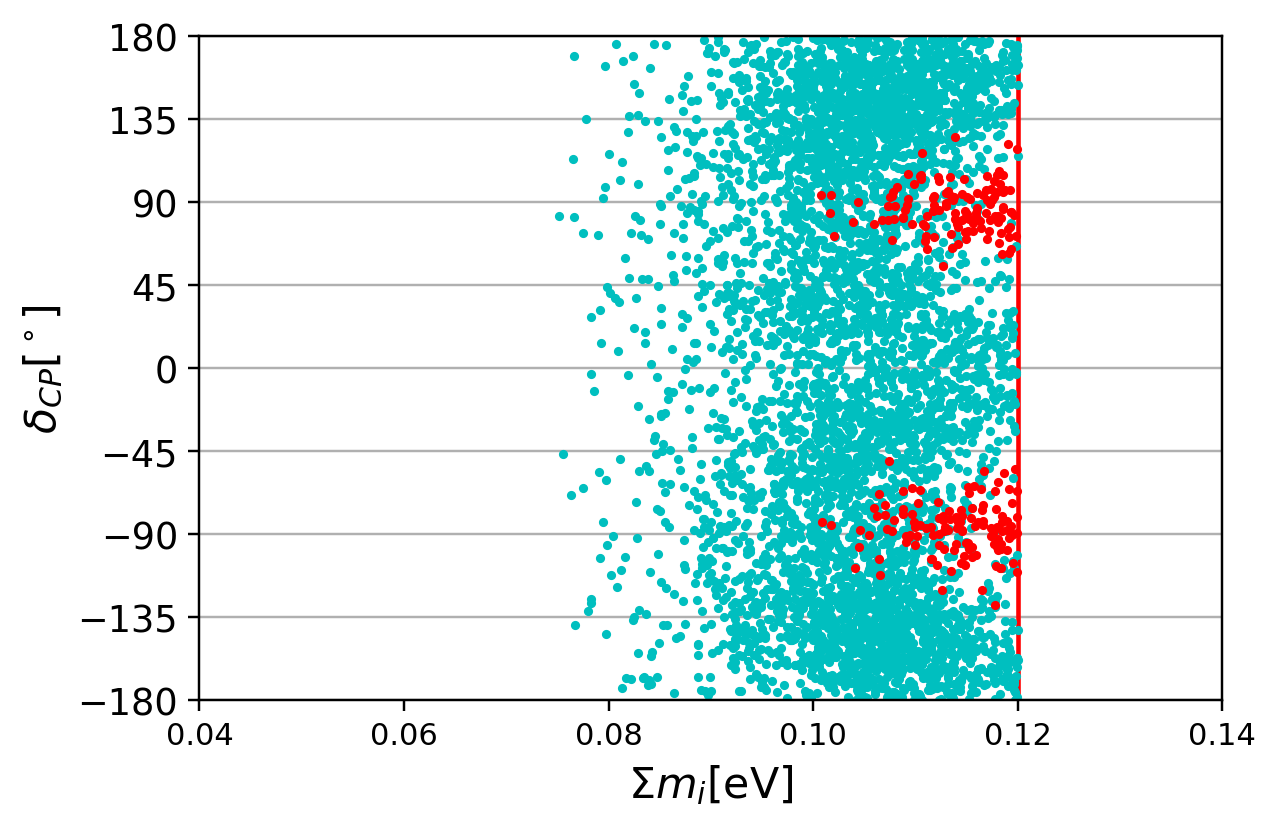}                                                                   
		\caption{The prediction of the neutrino mass sum and $\delta_{CP}$, where                                                  
		cyan  and red points denote NH and IH cases,  respectively.                                                                     
		The  red line represents the cosmological bound.}                                                             
		\label{fig:sum-cp}                                                                                                              
		\end{minipage}                                                                                                                  
		\phantom{=}                                                                                                                     
		\begin{minipage}{0.475\hsize}                                                                                                   
		\includegraphics[bb=0 0 400 280,width=\linewidth]{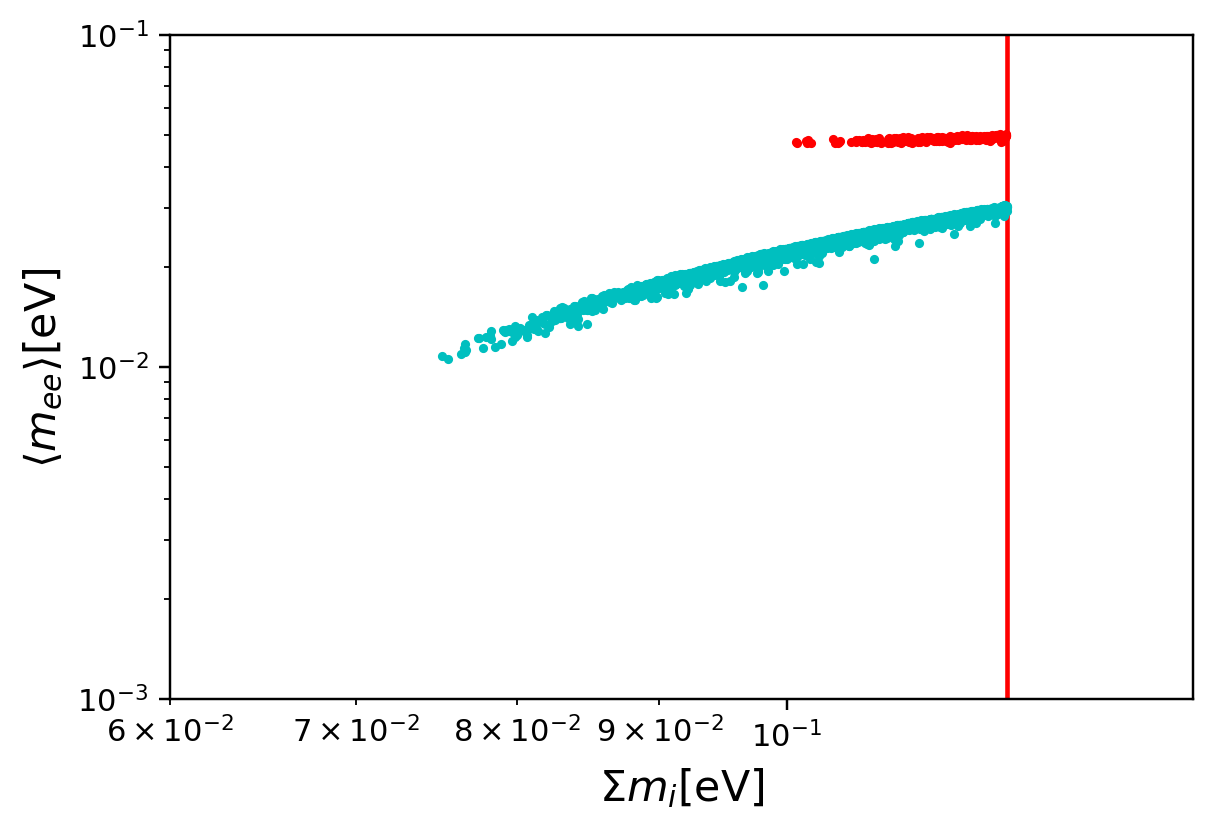}                                                                  
		\caption{The prediction of the effective mass for $0\nu\beta\beta$ decay, where                                                 
		cyan  and red points denote NH and IH cases,  respectively. The  red line represents the cosmological bound.} 
		\label{fig:mee}                                                                                                                 
		\end{minipage}                                                                                                                  
	\end{tabular}
\end{figure}
% --------------------------- %

%%%%%%%%%%%%%%%%%%%%%%%%%%%%%%%%%%%%%%%%%%%%%%%%%%%%%%%%%%%%%%%%%%%%%

Let us discuss our prediction of 
the leptonic CP phase, the effective mass of the neutrinoless double beta decay with the sum of neutrino masses.
We show the allowed region in the plane of
the sum of neutrino masses $\Sigma m_i$ and $\delta_{CP}$  in Fig.~\ref{fig:sum-cp}, where
cyan points and red points denote the predictions for NH and IH cases,  respectively.
For NH, the predicted Dirac CP violating phase is allowed in all range of
$[-180^\circ, 180^\circ]$ while  $\Sigma m_i$ is larger than  $75$ [meV].
In the case of IH,  $\delta_{CP}$ is predicted in the region of 
$\pm [50^\circ, 130^\circ]$.
Especially,  it is around $\pm 90^\circ$ near the lower bound of our prediction of the sum of neutrino masses, $100$ [meV].
It is noted that the lightest neutrino mass is larger than  $m_3 =1.61$ [meV] for IH.
The future development of neutrino oscillation experiments or cosmological analysis for the neutrino mass therefore is expected to test our model.

We also show a prediction of $\langle m_{ee}\rangle $ in the neutrinoless double beta decay in Fig.~\ref{fig:mee}.
The predicted region of the effective mass is about
$10< \langle m_{ee}\rangle < 30$ [meV] for NH and
$47 < \langle m_{ee}\rangle < 50$ [meV] for IH.
If the neutrinos are Majorana particle, the experiments for the neutrinoless double beta decay may test this model in the future.

%The predicted lower bound of the lightest neutrino mass is
%$m_1 = 10.4$ [meV] for NH.
%The predicted Majorana CP violating phases are shown in Fig.~\ref{fig:majo}.

%%%%%%%%%%%%%%%%%%%%%%%%%%%%%%%%%%%%%%%%%%%%
\begin{wrapfigure}[15]{r}[0pt]{8 cm}
	\vspace{0mm}
	\includegraphics[bb=0 0 400 280,width=\linewidth]{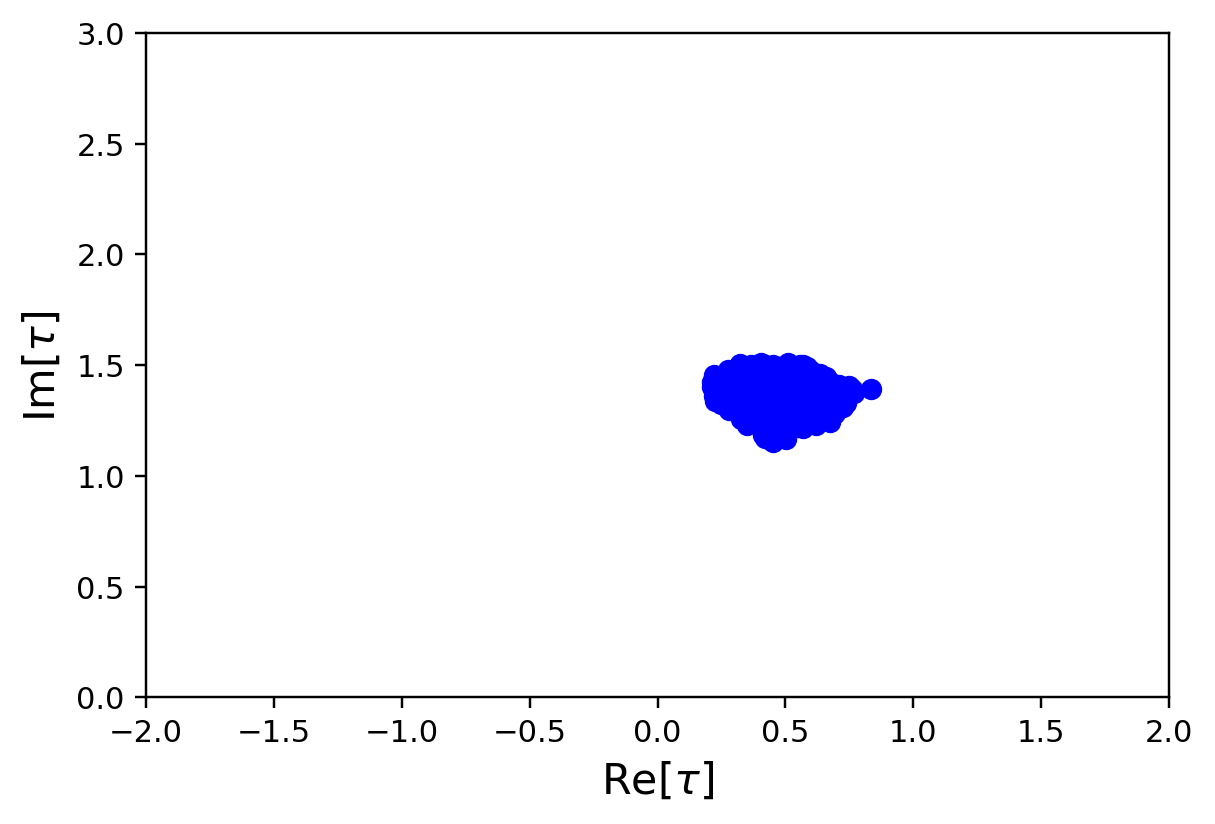}
	\caption{Allowed region of $\tau$ constrained 
		only from the quark sector.
	}
	\label{fig:tauquark}
\end{wrapfigure}
%%%%%%%%%%%%%%%%%%%%%%%%%%%%%%%%%%%%%%%%%%%%%

\subsection{Common modulus $\tau$ in quarks and leptons}

%%%%%%%%%%%%%%%%%%%%%%%%%%%%%%%%%%%%%%%%%%%%
\begin{figure}[b!]
	\begin{tabular}{ccc}
		\begin{minipage}{0.475\hsize}                                                                      
			\includegraphics[bb=0 0 400 280,width=\linewidth]{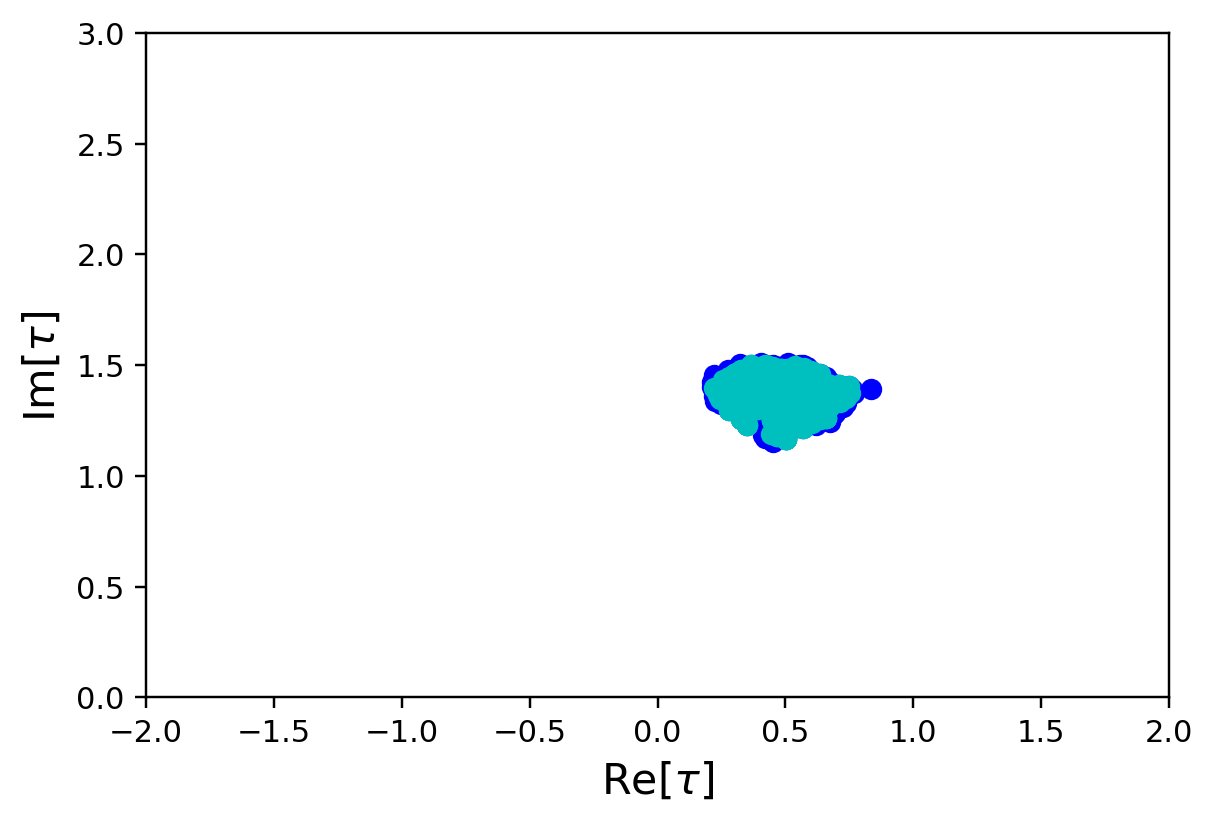}                              
			\caption{Allowed region of $\tau$ in both  quarks and leptons  (cyan points) for NH. 
				The region in quarks only is denoted by blue points.}                                              
			\label{fig:tauNH}                                                                                  
		\end{minipage}                                                                                     
		\phantom{=}                                                                                        
		\begin{minipage}{0.475\hsize}                                                                      
			\includegraphics[bb=0 0 400 280,width=\linewidth]{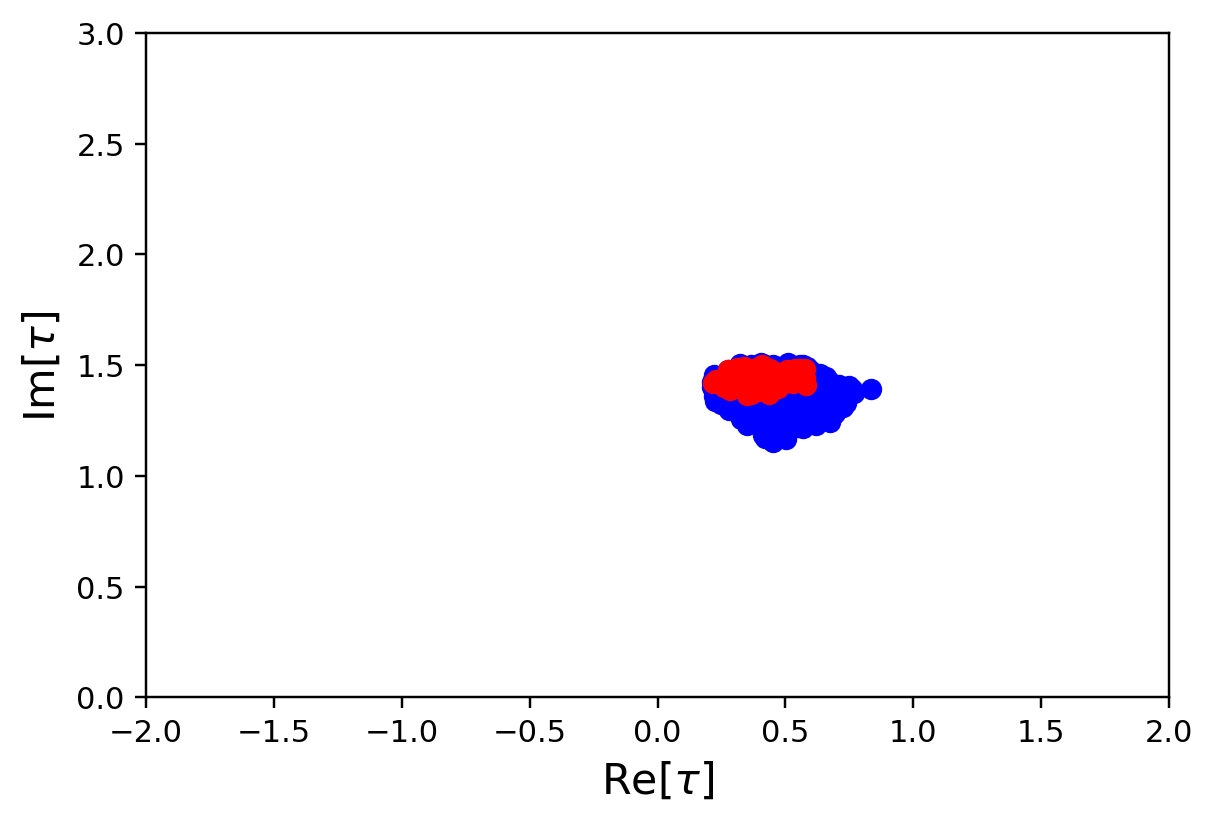}                              
			\caption{Allowed region of $\tau$ in both  quarks and leptons (red points) for IH.   
				The region in quarks only  is denoted by blue points. }                                            
			\label{fig:tauIH}                                                                                  
		\end{minipage}                                                                                     
	\end{tabular}
\end{figure}
The modulus $\tau$ is a key parameter to unify quark and lepton flavors.
We show allowed region of the modulus $\tau$ in Fig.~\ref{fig:tauquark} which leads to  successful  quark masses  and CKM mixing parameters at the GUT scale
within the $1\sigma$ range.
Both real and imaginary parts of $\tau$ are rather broad
as ${\rm Re} [\tau]=0.2$--$0.9$ and ${\rm Im} [\tau]=1.1$--$1.5$.

The quark and lepton mass matrices should have  the common modulus $\tau$.
Inputting  $\tau$ which is obtained in the quark sector  as well as other parameters of the quarks,
we have obtained the allowed region of $\tau$ which satisfies the observed $1\sigma$ range of the charged lepton mass ratios at the GUT scale and $3\sigma$ range of the PMNS parameters.
It is noted that there is no clear correlations between CKM and PMNS mixing parameters
because of the large number of free parameters embedded in our model. 

For the NH case, we plot the allowed region of  $\tau$ 
in Fig.~\ref{fig:tauNH}, where  $\tau$ of quark sector
is also shown.
Both regions are  almost overlapped.
However, for the IH case, 
the allowed region of $\tau$ is different from the one in the case of 
quarks only as seen   in Fig.~\ref{fig:tauIH}.
It is remarked that the allowed region clearly reduced compared with the one in quarks only.
 
Thus, we obtain the restricted common $\tau$ 
in spite of  many free parameters of our model.

\subsection{Proton decay}
We give a brief comment on the proton decay.
An SU(5) GUT model includes the color-triplet Higgs multiplets, which 
can lead to the proton decay \cite{Hisano:1992jj,Lucas:1996bc,Goto:1998qg,Murayama:2001ur,Hisano:2013exa}.
The color-triplet Higgs multiplets induce the following effective superpotential, 
\begin{equation}
	w_5 = \frac{1}{M_{H_c}} f_{u_i}e^{i\phi_i}f_{d_\ell}V^*_{k \ell}\varepsilon_{\alpha \beta \gamma}u^c_{i \alpha}e^c_iu^c_{k\beta} d^c_{\ell \gamma}
	\label{color-triplet}
\end{equation}
as well as $QQQL/M_{H_c}$ including quark doublet superfields $Q$, 
where $M_{H_c}$ is the mass of the color-triplet Higgs multiplets, 
$f_{u_i}$ and $f_{d_\ell}$ are Yukawa couplings of up-sector and down-sector quarks, and 
$\phi_i$ are their phases.\footnote{We follow the notation in \cite{Hisano:1992jj,Hisano:2013exa}.}
The above operator leads to the proton decay $p \rightarrow K^+ \nu$ through the higgsino exchange.
The factors $f_{u_3}f_{d_\ell}V^*_{1 \ell}$ with $\ell = 1,2$ are crucial to estimate the proton life-time, 
because the couplings among $u_R$, $d_R$($s_R$), right-handed stop and right-handed stau are important in this process.
Then, the proton life-time is given as \cite{Hisano:2013exa}:
\begin{align}
	\tau_P\simeq 4\times 10^{35}\times \sin^4 2\beta                      
	\left ( \frac{0.1}{\hat A_R}  \right )^2                              
	\left ( \frac{M_S}{100{\rm TeV}}  \right )^2                          
	\left ( \frac{M_{H_c}}{10^{16}{\rm GeV}}  \right )^2  \ {\rm yrs} \ , 
\end{align}
where $\hat A_R$ is the renormalization factor and 
$M_S$ is the sfermion mass scale.
The proton life-time is longer than  the observed lower bound $10^{33}$ yrs
\cite{Tanabashi:2018oca} 
for the case of   $M_{H_c}\simeq 2\times 10^{16}$GeV
if  $M_S\geq 10$TeV and   $\tan\beta\leq 3$.
Since 
 our numerical results of quark/lepton mass matrices
 are changed only within a few percent  in the range of 
 $\tan\beta= 3$--$10$ as stated in footnote 1,
  $M_S=10$TeV  is the minimal one, which is consistent with our numerical results of the quark/lepton flavor mixing,  to protect the proton decay
  \footnote{If $\tan\beta=10$ is taken,  $M_S$ should be larger than 
  	$100$TeV.}.

We may have additional contributions to the effective potential in Eq.(\ref{color-triplet}), because 
our cut-off scale $\Lambda$ is slightly higher than the GUT scale,  $\langle H_{24} \rangle /\Lambda =0.3$.
For example, the following term is allowed by the symmetries,
\begin{equation}
	w'_5 = \frac{f}{\Lambda} T_3 T_3 T_3 F_3,
\end{equation}V
where $f$ is a modulus-independent coupling constant.
The field $T_3$ includes $u_R$ by the factor $c^u_{31}Y_2 \sim 5 \times 10^{-3}$, while 
the field $F_3$ includes $d_R$ and $s_R$ by the factors  $c^d_{31}Y_2 \sim 1 \times 10^{-2}$ and  $c^d_{31}Y_1 \sim 1 \times 10^{-1}$.
Thus, the above operator leads to the couplings among $u_R$, $d_R$($s_R$), right-handed stop and right-handed stau 
with a similar suppression or strong suppression compared with $f_{u_3}f_{d_\ell}V^*_{1 \ell} = {\cal O}(10^{-4})$ for $\ell = 1,2$, 
when $f={\cal O}(1)$.
In our model we set  $\langle H_{24} \rangle /\Lambda =0.3$ and $\langle H_{24} \rangle \simeq 2 \times 10^{16}$ GeV.
For $M_{H_c} < \Lambda$ and $f \lesssim 1$,  the processes including the color-triplet Higgs multiplets of Eq.(\ref{color-triplet}) would be 
dominant in the proton decay.
Similarly, we can discuss the operators including $T_{1,2}$ and $F_{1,2}$ although they should have modulus-dependent 
couplings.

%%%%%%%%%%%%%%%%%%%%%%%%%%%%%%%%%%%%%%%%%%%%%%%%%%%%%%%%%%%%%%%%%%%
%%%%%%%%%%%%%%%%%%%%%%%%%%%%%%%%%%%%%%%%%%%%%%%%%%%%%%%%%%%%%%%%%%%
%%%%%%%%%%%%%%%%%%%%%%%%%%%%%%%%%%%%%%%%%%%%%%%%%%%%%%%%%%%%%%%%%%%
\section{Summary and discussions}

We have presented a flavor model with the $S_3$ modular invariance in the framework of SU(5) GUT and discussed the CKM and PMNS mixing parameters of both quark and lepton sectors, respectively.
We have considered six-dimensional compact space $X^6$ in addition to our four-dimensional space-time and
supposed that the six-dimensional compact space has some constituent parts and they include a two-dimensional compact space $X^2$.
Then,  the quarks and leptons have the same modular symmetry $S_3$ and the same value of $\tau$ in our setup.
We note that our model does not require any gauge singlet scalars such as flavons.
The difference between mass eigenvalues of down-type quarks and charged leptons is realized by the 24-dimensional adjoint Higgs multiplet $H_{24}$.

The setup of our model is
reasonably accepted in the neutrino phenomenology 
if the  right-handed neutrino masses are taken to be around $10^{15}$GeV.
Their favored range  are fairly limited; masses larger than $10^{15}$ GeV are basically disfavored by perturbativity of the neutrino Yukawa couplings, whereas lower masses require a more significant suppression in the dimension-five operators, or more severe cancellation between the operators and the bare mass terms.

We have analyzed our model numerically and  found 
parameter regions which are consistent with both the observed CKM and PMNS mixing parameters for both  NH and IH cases.
The predicted  Dirac CP violating phase is allowed
in all range of $[-180^\circ, 180^\circ]$ for NH.
The sum of neutrino masses is larger than  $75$ [meV].
For IH, it is predicted in the region of 
$\pm [50^\circ, 130^\circ]$.
Especially,  it is around $\pm 90^\circ$ near the lower bound of our prediction of the sum of neutrino masses, $100$ [meV].
It is expected to test our model by the astronomical observation for the neutrino mass constraint as well as the precise observation for the Dirac CP violating phase. 

We have also predicted the effective mass in the neutrinoless double beta decay $\langle m_{ee}\rangle$,
which is 
$10< \langle m_{ee}\rangle < 30$ [meV] for NH and
$47 < \langle m_{ee}\rangle < 50$ [meV] for IH.
The development of the experiments for the neutrinoless double beta decay is also expected to test our model.
It is also noted that the proton life-time is enough long
compared with the observed lower bound $10^{33}$ yrs.

Since our model has a large number of free parameters,
distinct correlations between CKM and PMNS mixing parameters are not found.
However, the common value of  modulus $\tau$ is clearly obtained 
by imposing the experimental data of CKM and PMNS mixing parameters as well as quark and lepton masses.
If we can build a flavor model with a small number of free parameters in a specific GUT framework,
it is expected to find  correlations between the CKM  and  PMNS matrices.
 
% --------------------------------------------------------------------- %
% --------------------------------------------------------------------- %
% --------------------------------------------------------------------- %

%-------- acknowledgement -------%
\vspace{0.5cm}
\noindent

{\bf Acknowledgement}\\

This work is supported by  MEXT KAKENHI Grant Number JP19H04605 (TK), and 
JSPS Grants-in-Aid for Scientific Research 18J11233 (THT).
The work of YS is supported by JSPS KAKENHI Grant Number JP17K05418 and Fujyukai Foundation.
%-------- Appendix -------%

\appendix
\section*{Appendix}

\section{Modular forms of $S_3$ modular group}
\label{sec:S_3 modular forms}
The Dedekind eta-function $\eta(\tau)$ is defined by 
\begin{align}
	\eta(\tau) = q^{1/24} \prod_{n =1}^\infty (1-q^n)~, 
\end{align}
where $q = e^{2 \pi i \tau}$.
By use of $\eta(\tau)$, 
the modular forms of weight 2 corresponding to the $S_3$ doublet are written by \cite{Kobayashi:2018vbk},
\begin{eqnarray}
	\label{eq:Y-S3}
	Y_1(\tau) &=& \frac{i}{4\pi}\left( \frac{\eta'(\tau/2)}{\eta(\tau/2)}  +\frac{\eta'((\tau +1)/2)}{\eta((\tau+1)/2)}
	- \frac{8\eta'(2\tau)}{\eta(2\tau)}  \right), \nonumber \\
	Y_2(\tau) &=& \frac{\sqrt{3}i}{4\pi}\left( \frac{\eta'(\tau/2)}{\eta(\tau/2)}  -\frac{\eta'((\tau +1)/2)}{\eta((\tau+1)/2)}   \right) , \label{doubletY}  \nonumber
\end{eqnarray}
where we use the following  basis of $S_3$ generators
$S$ and $T$ in the doublet representation:
\begin{equation}
	S = \frac{1}{2}\left(
	\begin{array}{cc}
		-1        & -\sqrt{3} \\
		-\sqrt{3} & 1         
	\end{array}\right), \qquad\qquad
	T = \left(
	\begin{array}{cc}
		1 & 0  \\
		0 & -1 
	\end{array}\right).
	\label{S3base}
\end{equation}
The doublet modular forms have the following  $q$-expansions:
\begin{align}
	Y^{(2)}_{\bf 2}=\begin{pmatrix}Y_1(\tau)                      \\Y_2(\tau)
	\end{pmatrix}_{\bf 2} =                                       
	\begin{pmatrix}                                               
	\frac{1}{8}+3q+3q^2+12q^3+3q^4+\dots                          \\
	\sqrt{3}q^{1/2}(1+4q+6q^2+8q^3+\dots ) \end{pmatrix}_{\bf 2}. 
\end{align}
Since we work in the basis of Eq.(\ref{S3base}),
the tensor product of two doublets is expanded by
\begin{eqnarray}
	\left(
	\begin{array}{c}
		x_1 \\ x_2
	\end{array}\right)_{\bf 2} \otimes
	\left(
	\begin{array}{c}
		y_1 \\ y_2
	\end{array}\right)_{\bf 2} &=&\left(x_1y_1+x_2y_2\right)_{\bf 1}
	\oplus\left(x_1y_2-x_2y_1\right)_{\bf 1'}
	\oplus\left(
	\begin{array}{c}
		x_1 y_1-x_2 y_2 \\ -x_1 y_2- x_2 y_1
	\end{array}\right)_{\bf 2}.
\end{eqnarray}
By using the tensor product of the two doublets $(Y_1(\tau), Y_2(\tau))^T$, we can construct modular forms of weight 4,  $Y^{(4)}$:
\begin{align}
	{\bf 1}~ & : ~~ Y^{(4)}_{\bf 1} = \left(Y_1(\tau)^2+Y_2(\tau)^2\right)_{\bf 1} , \\
	{\bf 2}~ & : ~~ Y^{(4)}_{\bf 2} =                                                
	\begin{pmatrix}
	Y_1(\tau)^2 - Y_2(\tau)^2  \\
	-2Y_1(\tau)Y_2(\tau)
	\end{pmatrix}_{\bf 2} .
\end{align}
The $S_3$ singlet ${\bf 1}'$ modular form of the weight 4 vanishes.

\section{Lepton mixing matrix }

Supposing neutrinos to be Majorana particles, 
the PMNS matrix 
is parametrized in terms of the three mixing angles $\theta _{ij}$ $(i,j=1,2,3;~i<j)$,
one CP violating Dirac phase $\delta _\text{CP}$ and two Majorana phases 
$\alpha_{21}$, $\alpha_{31}$  as follows \cite{Tanabashi:2018oca}:
\begin{align}
	U_\text{PMNS} =
	\begin{pmatrix}
	c_{12} c_{13} & s_{12} c_{13}              & s_{13}e^{-i\delta_\text{CP}} \\
	-s_{12} c_{23} - c_{12} s_{23} s_{13}e^{i\delta_\text{CP}} &
	c_{12} c_{23} - s_{12} s_{23} s_{13}e^{i\delta_\text{CP}} & s_{23} c_{13} \\
	s_{12} s_{23} - c_{12} c_{23} s_{13}e^{i\delta_\text{CP}} &
	-c_{12} s_{23} - s_{12} c_{23} s_{13}e^{i\delta_\text{CP}} & c_{23} c_{13}
	\end{pmatrix}
	%\times
	\begin{pmatrix}
	1             & 0                          & 0                            \\
	0             & e^{i\frac{\alpha_{21}}{2}} & 0                            \\
	0             & 0                          & e^{i\frac{\alpha_{31}}{2}}   
	\end{pmatrix},
	\label{UPMNS}
\end{align}
where $c_{ij}$ and $s_{ij}$ denote $\cos\theta_{ij}$ and $\sin\theta_{ij}$, respectively.

In terms of these parametrization
and three neutrino masses, 
the effective mass in the neutrinoless double beta decay is given as follows:
\begin{align}
	\langle m_{ee}	\rangle=\left| m_1 c_{12}^2 c_{13}^2+ m_2s_{12}^2 c_{13}^2 e^{i\alpha_{21}}+ 
	m_3 s_{13}^2 e^{i(\alpha_{31}-2\delta_{CP})}\right|  \ .                                    
\end{align}

% ------------------ %
% --- References ---- %
% ------------------ %
%\newpage

\end{document}